\documentstyle[aps,preprint]{revtex}
\tightenlines
\newcommand{\p}[1]{(\ref{#1})}
\def\beq{\begin{eqnarray}} \def\eeq{\end{eqnarray}}
\def\beqstar{\begin{eqnarray*}} \def\eeqstar{\end{eqnarray*}}
\draft
\begin {document}

\title{Fermi--liquid theory of \protect\\ superfluid
asymmetric nuclear matter }

\author{\fbox{A. I. Akhiezer},  A. A. Isayev,
 S. V. Peletminsky,   A. A.  Yatsenko}

\address{Kharkov Institute of Physics and Technology,
 Kharkov, 61108, Ukraine}
\maketitle

\begin{abstract}
      Influence of asymmetry on
superfluidity  of nuclear matter with triplet--singlet pairing of
nucleons (in  spin and isospin spaces) is considered
within the framework of a
Fermi--liquid theory. Solutions of self--consistent equations for the
critical temperature and the energy gap at $T=0$ are obtained with the use
of  Skyrme effective nucleon interaction.  It is shown, that if the
chemical potentials of protons and neutrons are determined in the
approximation of ideal Fermi--gas, then the energy gap for some values of
density and asymmetry parameter of nuclear matter demonstrates double--valued
behavior.  However, with account for the feedback of pairing correlations
through the normal distribution functions of nucleons two--valued behavior of
the energy gap turns into universal one--valued behavior. At $T=0$
superfluidity arises and disappears as a result of a first order phase
transition in density.  \end{abstract} 

\narrowtext

It is well established, that
neutron--proton ($np$) pairing plays an essential role in the description of
superfluidity of finite nuclei with $N=Z$ (see Refs.~\cite{G,RSS} and
 references therein) and symmetric nuclear matter \cite{AFR}--\cite{GSM}.
In astrophysical context  $np$ pairing correlations         can be
important for the description of $r$--process \cite{KBT,CDK} and cooling of
neutron stars, which permit pion  or kaon condensation \cite{BB,T}. In this
Rapid Communication we shall investigate the influence of asymmetry on $np$
superfluidity of nuclear matter.  Previously this problem was treated with
the use of various approaches and potentials of NN interaction.  In
particular, the cases of $^3S_1$--$^3D_1$ and $^3D_2$ pairing  were
considered in Refs.~\cite{AFR,ARS} on the basis of the Thouless criterion for
the thermodynamic $T$ matrix. As a potential of NN interaction,   the
Graz~II and Paris potentials were chosen, respectively.  Superfluidity in
$^3S_1$--$^3D_1$ pairing channel was studied also   in Ref.~\cite{SAL} within
BCS theory of superconductivity with the use of the Paris potential in the
separable form.  Investigations, based on the Thouless criterion, deduce the
suppression of $np$ pairing correlations with increase of isospin asymmetry.
However, the Thouless criterion can be exploited for finding the critical
temperature only, but does not permit one to draw any conclusions about
superfluidity with a finite gap. The studies in Ref.~\cite{SAL}, based on the
BCS theory, were carried out with the use of the bare interaction and the
single particle spectrum of a free Fermi gas and give, thus, overestimated
values of the energy gap. The effect of ladder--renormalized single particle
spectrum~\cite{ZBL} on the magnitude of the energy gap in $^3S_1$--$^3D_1$
pairing channel was investigated in Ref.~\cite{SL}.    The Argonne $V_{14}$
potential was explored as input for determination of the single particle
energy and  the bare interaction in the form of the Paris potential was used
to evaluate the energy gap.  The use of the bare interaction in the gap
equation  seems to be  a very strong simplification, because medium
polarization strongly reduces the magnitude of the gap (see Ref.~\cite{SCL}
for the influence of the polarization effects on the pairing force in the
$^1S_0$ channel). In principle, the effective pairing interaction should be
obtained by means of Brueckner renormalization, which gives the correct
interaction after modifying the bare interaction for the effect of nuclear
medium.  However, the issue of a microscopic many--body calculation of the
effective pairing potential is a complex one and still is not solved. For
this reason, it is quite a natural step to develop some kind of a
phenomenological theory, where instead of microscopical calculation of the
pairing interaction one exploits the phenomenological effective interaction.
We shall investigate the influence of asymmetry on superfluid properties of
nuclear matter, using Landau's semiphenomenological theory of a Fermi--liquid
(FL).  In the Fermi--liquid model the normal and anomalous FL interaction
amplitudes are taken into account on an equal footing. This will allow us to
consider consistently the influence of the  FL amplitudes on superfluid
properties of nuclear matter.  Besides, as a potential of NN interaction we
choose the Skyrme effective forces, describing the interaction of two
nucleons in the presence of nucleon medium. The Skyrme forces are widely used
in the description of nuclear system properties and, in particular, they were
exploited for the description of superfluid properties of finite nuclei
\cite{DFT,RDN} as well as infinite symmetric nuclear matter
\cite{SYK}-\cite{AIP2}.

  The basic formalism is laid out in more detail
in Ref. \cite{AIP}, where superfluidity of symmetric nuclear matter was
studied.
 As
shown there, superfluidity with triplet--singlet (TS)
pairing of nucleons (total spin $S$ and isospin $T$ of a pair are equal
$S=1$, $T=0$) is
realized near the saturation density in symmetric nuclear matter with the
Skyrme interaction.  Therefore,  we shall study further the influence of
asymmetry on superfluid properties of TS phase of nuclear matter. For the
states with the projections of total spin and isospin $S_z=T_z=0$ the normal
distribution function $f$ and the anomalous distribution function $g$ have
the form \begin{equation} f( {\bf p})=f_{00}({\bf p})\sigma_0\tau_0+
f_{03}({\bf p})\sigma_0\tau_3,\quad g({\bf p})=g_{30}({\bf
p})\sigma_3\sigma_2\tau_2\, \label{1}\end{equation} where $\sigma_i,\tau_k$
are the Pauli matrices in spin and isospin spaces.  The operator of
quasiparticle energy $\varepsilon$ and    the matrix order parameter $\Delta$
of the system for the energy functional, being invariant with respect to
rotations in spin and isospin spaces, have the analogous structure
\begin{equation} \varepsilon({\bf p})=\varepsilon_{00}({\bf p})\sigma_0\tau_0+
\varepsilon_{03}({\bf p})\sigma_0\tau_3,\quad
\Delta({\bf p})=\Delta_{30}({\bf p})\sigma_3\sigma_2\tau_2.
\label{5}\end{equation}
Using the minimum principle of the thermodynamic potential and procedure of
  block diagonalization \cite{AKP},  one can express evidently
the distribution functions $f_{00},
f_{03}$ and $g_{30}$ in terms of the quantities $\varepsilon$ and $\Delta$:
\beq
f_{00}&=&\frac{1}{2}-\frac{\xi_{00}}{4E}\Bigl(\tanh\frac{E+\xi_{03}}{2T}
+\tanh\frac{E-\xi_{03}}{2T}\Bigr),  \label{6'}\\
f_{03}&=&-\frac{1}{4}\Bigl(\tanh\frac{E+\xi_{03}}{2T}
-\tanh\frac{E-\xi_{03}}{2T}\Bigr),\label{6}\\
g_{30}&=&-\frac{\Delta_{30}}{4E}\Bigl(\tanh\frac{E+\xi_{03}}{2T}
+\tanh\frac{E-\xi_{03}}{2T}\Bigr).\label{6''}\eeq
Here
\beqstar
E=\sqrt{\xi_{00}^2+\Delta_{30}^2},\;&\xi_{00}=\varepsilon_{00}-\mu_{00}^0,\;
&\xi_{03}=\varepsilon_{03}-\mu_{03}^0,\\
\mu_{00}^0=&\displaystyle{\frac{\mu_p^0+\mu_n^0}{2}},\;
\mu_{03}^0=&\displaystyle{\frac{\mu_p^0-\mu_n^0}{2}},
\eeqstar
$T$ is temperature, $\mu_p^0$ and $\mu_n^0$ are chemical potentials of
protons and neutrons.  To obtain the closed system of  equations for the
quasiparticle energy $\varepsilon$ and the energy gap $\Delta$, it is
necessary to express the quantities $\varepsilon,\Delta$ through the
distribution functions $f$ and $g$. For this purpose one has to set the
energy functional ${\cal E}(f,g)$ of the system. In the case
of asymmetric nuclear matter with TS pairing of nucleons
the energy functional is characterized by two normal $U_0,U_2$ and
one anomalous $V_1$ FL amplitudes \cite{AIP}. Differentiating the functional
${\cal E}(f,g)$ with respect to $g$ \cite{AKP} and using Eq.~\p{6''}, one can
obtain the gap equation in the form \beq \Delta_{30}({\bf
p})&=&-\frac{1}{4\cal V}\sum_{{\bf q}}V_1({\bf p},{\bf
q})\frac{\Delta_{30}({\bf q})}{E({\bf q})}\label{15} \\&&
\times\Bigl\{\tanh\frac{E({\bf q})+\xi_{03}({\bf q})}{2T} +\tanh\frac{E({\bf
q})-\xi_{03}({\bf q})}{2T}\Bigr\}. \nonumber \eeq The anomalous interaction
amplitude $V_1$ in the Skyrme model reads \cite{AIP} \beq V_{1}({\bf p},{\bf
q})&=&t_0(1+ x_0)+\frac{1}{6}t_3\varrho^\beta(1+x_3)\label{16}\\
&&+\frac{1}{2\hbar^2}t_1(1+
x_1)({\bf p}^{\,2}+{\bf q}^{\,2}), \nonumber\eeq
where $\varrho$ is  density of nuclear matter,
$t_i,x_i,\beta$      are
some phenomenological
parameters, which differ for
various versions of the Skyrme forces (later we shall use the SkP
potential \cite{DFT}).
 Equation~\p{15} should be solved jointly
with equations \beq \frac{1}{\cal V}\sum_{{\bf p}}
\Bigl\{2-\frac{\xi_{00}({\bf p})}{E({\bf
p})}&                   \nonumber         \\
\times\Bigl(\tanh\frac{E({\bf p})+\xi_{03}({\bf p})}{2T}
&+\tanh\displaystyle{\frac{E({\bf p})-\xi_{03}({\bf p})}{2T}}\Bigr)\Bigr\}
=\varrho,\label{15'}
 \eeq
\begin{equation}
\frac{1}{\cal V}\sum_{{\bf p}}
\Bigl\{\tanh\frac{E({\bf p})+\xi_{03}({\bf p})}{2T}
-\tanh\frac{E({\bf p})-\xi_{03}({\bf p})}{2T}\Bigr\}
=\alpha\varrho,\label{15''}
 \end{equation}
being the normalization conditions for the normal distribution functions
$f_{00}, f_{03}$. In Eq.~\p{15''} the quantity
 $\alpha=(\varrho_n-\varrho_p)/\varrho$ is
the asymmetry parameter of nuclear matter,
$\varrho_p,\varrho_n$ are the partial number densities of
protons and neutrons. Note that the account of the normal FL amplitudes in the
case of the effective Skyrme interaction, being quadratic in momenta, is
reduced to the renormalization of free nucleon masses and chemical potentials.
Expressions for the quantities $\xi_{00}, \xi_{03}$, which enter in
Eqs.~\p{15},\p{15'},\p{15''}, with regard for the explicit form of the
amplitudes $U_0,U_2$ \cite{AIP} read
$$\xi_{00}=\frac{p^2}{2m_{00}}-\mu_{00},\;\xi_{03}=\frac{p^2}{2m_{03}}
-\mu_{03},           $$
where the effective nucleon mass $m_{00}$ and  effective isovector mass
  $m_{03}$ are defined by
 the formulas:
\beq
\frac{\hbar^2}{2m_{00}}&=&\frac{\hbar^2}{2m_{00}^0}+\frac{\varrho}{16}
[3t_1+t_2(5+4x_2)],\label{18}\\
\frac{\hbar^2}{2m_{03}}&=&
\frac{\alpha\varrho}{16}[t_1(1+2x_1)-
t_2(1+2x_2)],\nonumber\eeq
 $m_{00}^0$ being the bare mass of a nucleon. The renormalized chemical
potentials $\mu_{00},\mu_{03}$ should be determined from Eqs.  \p{15'},
 \p{15''} and in the leading approximation on the ratios
$T/\varepsilon_F,\Delta/\varepsilon_F$ have the form \begin{equation}
\mu_{00}=\frac{1}{2}(\mu_p+\mu_n),\;
\mu_{03}=\frac{1}{2}(\mu_p-\mu_n); \quad
\mu_{p,n}=\frac{\hbar^2k^2_{F_{p,n}}}{2m_{p,n}},\label{20}\end{equation}
where $k_{F_{p,n}}=(3\pi^2\varrho_{p,n})^{1/3}$, $m_p$ and $m_n$ are the
proton and neutron effective masses, defined                 as
$$\frac{2}{m_{00}}=\frac{1}{m_p}+\frac{1}{m_n},\quad
\frac{2}{m_{03}}=\frac{1}{m_p}-\frac{1}{m_n}.$$

The critical temperature of transition to TS
superfluid phase is found from Eq.~\p{15}, determining the energy gap, in the
linear on $\Delta$ approximation.
Considering, that the interaction amplitude $V_1$ is not equal to zero only
in a narrow layer near the Fermi--surface, $|\xi_{00}|\leq\theta$ (we shall
set $\theta=0.1\mu_{00}$) and entering new variables $x=\xi_{00}/\mu_{00}$,
$\theta_0=\theta/\mu_{00}$, $T_\mu=T/\mu_{00}$,  we present this equation in
the form
\beq
1&=&\frac{g}{4}\int_{-\theta_0}^{\theta_0}\frac{dx}{x}\label{24}\\
&&\times
\Bigl\{\tanh\frac{x[1+\frac{m_{00}}{m_{03}}]+\psi}{2T_\mu}+
\tanh\frac{x[1-\frac{m_{00}}{m_{03}}]-\psi}{2T_\mu}\Bigr\},\nonumber
\eeq
\beqstar
g = -\nu_FV_1(p=p_F,q&=&p_F),\;\;\psi = \frac{m_{00}}{m_{03}}
-\frac{\mu_{03}}{\mu_{00}},\\
\nu_F = \frac{m_{00}p_F}{2\pi^2\hbar^3},\;\;&p_F& = \sqrt{2m_{00}\mu_{00}}
\eeqstar

The results of numerical integration of Eq.~\p{24} are shown in
Fig.\ \ref{fig1}.
For small values of
asymmetry $\alpha$ there exist such regions of large and low densities of
nuclear matter, for which we have two critical temperatures.
When $\alpha$ increases, these regions begin to approach and at
some value $\alpha=\alpha_c$ ($\alpha_c\approx0.071$) it takes
place contiguity of the regions, so that we have always only two critical
temperatures (for the regions, where solutions exist). For $\alpha>\alpha_c$
the phase curves are separated from the density axis and turn into the closed
oval curves.  Under further increase of $\alpha$ the dimension of the oval
curves is reduced and at some $\alpha=\alpha_m$ the oval curves contract to a
point ($\alpha_m\approx0.092 $).
  For the values $\alpha>\alpha_m$ the triplet--singlet
superfluidity fails.
Note, that our results concerning two--valued behavior of the critical
temperature qualitatively agree with the results of Ref.~\cite{ARS}, where
$^3D_2$ pairing of nucleons with the Paris NN potential was considered.

 As follows from Eq.~\p{15}, an equation determining
the energy  gap at $T=0$ has the form
\beq \Delta_{30}({\bf p})&=&-\frac{1}{4\cal V}
\sum_{{\bf q}}V_1({\bf p},{\bf q})
\frac{\Delta_{30}({\bf q})}{E({\bf q})}\label{25}\\
&&\times\left\{\mbox{sgn}
\,(E({\bf q})+\xi_{03}({\bf
q}))+\mbox{sgn}\,(E({\bf q})-\xi_{03}({\bf
q}))\right\} \nonumber\eeq
Passing in Eq.~\p{25} to integration on a layer,
we arrive at the equation for determining the dimensionless gap
$y=\Delta_{30}/\mu_{00}$:  \begin{equation}
1=\frac{g}{4}\int_{-\theta_0}^{\theta_0}\frac{dx}{\sqrt{x^2+y^2}}\Bigl\{
1+
\mbox{sgn}\,\bigl(\sqrt{x^2+y^2}-\frac{m_{00}}{m_{03}}x-\psi
\bigr)\Bigr\}
\label{26}\end{equation}
(for $\alpha>0$ it holds true
$m_{03}>0,\mu_{03}<0$).
The contribution to the integral gives the domain on $x$, for which
the function, standing as an
argument of the function $\mbox{sgn}$, is positive.
  In particular,  such values of the gap, density and asymmetry
parameter of nuclear matter are possible, that this function  has no
roots with respect to $x$.  In this case the whole domain from $-\theta_0$ to
$\theta_0$ gives the contribution to the integral in Eq.~\p{26} and we arrive
at the equation of the BCS type at $T=0$
  with the solution $y=\theta_0/(\sinh(1/g))$.

Let us  first  find the solutions of Eq.~\p{26} in the case when the chemical
potentials $\mu_{00},\mu_{03}$ are given in the main approximation on
$\Delta/\varepsilon_F$.  The results of numerical integration of
Eq.~\p{26} are presented in Fig.\ \ref{fig2}.
In the case of symmetric
nuclear matter ($\alpha=0$) we
obtain the phase curve with one--valued behavior of the gap.   For small values
of asymmetry $\alpha$ there
exist such regions of large and low densities of nuclear matter (excluding
some vicinity of the point $\varrho=0$), for which we have two values of the
energy gap, where one of these values is the solution of the BCS type  and
practically coincides with the corresponding value of the gap for the case
$\alpha=0$.  For $\alpha$ less than some $\alpha_c$, the energy gap
demonstrates double--valued  behavior in the intervals
$(\varrho_{min},\varrho'_{min})$ and $(\varrho_{max}',\varrho_{max})$, while
in the interval $(\varrho_{min}',\varrho_{max}')$ it has one--valued behavior
(see Table\ \ref{table1}  for the values of various boundary points
$\varrho$).
When $\alpha$ increases, the regions with double--valued
behavior of the gap begin to approach and at   $\alpha=\alpha_c$ (the same
$\alpha_c$ as for the phase curves $T_c(\varrho)$) it takes place contiguity
of the regions with two solutions.  For $\alpha>\alpha_c$ two branches of the
phase curves are separated from the density axis and combine to one curve,
beginning and ending in some points of the phase curve with $\alpha=0$.  In
this case the energy gap differs from zero only in the interval
$(\varrho_{min},\varrho_{max})$, where it has double--valued
behavior. When $\alpha$ increases further, the boundary points of the phase
curves move towards and at some $\alpha=\alpha_m'$ (not equal to $\alpha_m$
for the phase curves $T_c(\varrho)$) the branches of solutions contract to a
point.  The value $\alpha_m'$ determines the maximum value of the asymmetry
parameter, when TS superfluidity exists at $T=0$ ($\alpha_m'\approx0.179$).

Let us  now find the solutions of Eq.~\p{26} while accounting for the
influence of the finite size of the gap on the chemical potentials
$\mu_{00},\mu_{03}$. The results of integration of
Eq.~\p{26} in this case are presented in Fig.\ \ref{fig3}.
Here the solutions of
Eqs.~\p{26},   \p{15'}, and \p{15''} obtained for different $\alpha$,
correspond to the different parts of the dome--shaped curve, contained by the
dashed lines of the same type for a given $\alpha$.  One can see, that taking
into account  the feedback of the finite size of the gap through the normal
distribution functions $f_{00}, f_{03}$ in Eqs.~\p{6'}, \p{6} leads to the
qualitative change:  instead of two--valued behavior of the gap we have
universal one-valued behavior. The first solution of the BCS type, obtained
in uncoupled calculation, remains practically unchanged in self--consistent
treatment of the gap equation \p{26} and with sufficiently high accuracy
equals to its value at $\alpha=0$ in the self--consistent determination.  The
second solution in uncoupled scheme, to which corresponds the smaller gap
width, under simultaneous iterations of Eqs.~\p{15'}, \p{15''}, \p{26} tends
to the first solution of the BCS type.  Taking into account  the finiteness
of the gap results in the reduction of the threshold asymmetry, at which
superfluidity disappears, to the value $\alpha_m'\approx0.029$.  Thus, in
spite of the smallness of the ratio $\Delta/\varepsilon_F$
($\Delta/\varepsilon_F\leq0.12$ for all densities $\varrho$), the backward
influence of pairing correlations is significant.  This is explained by the
fact, that if the quantity $\Delta$ in Eqs.~\p{15'}, \p{15''} differs from
zero, then the absolute value of the chemical potential $\mu_{03}$ increases
 a few times as against its value at $\Delta=0$.  The increase of $|\mu_{03}|$
is equivalent to the increase of the effective shift between neutron and
proton Fermi surfaces, that leads to significant reduction of the threshold
asymmetry. As at $\alpha\not=0$ the gap is  finite everywhere, superfluidity
arises and disappears under changing density by means of a first order phase
transition. In principle, this phase transition can be observed in laboratory
conditions under the study of intermediate--energy heavy ion reactions. If we
assume that the final stage of the reaction can be described as an expansion
of a compound nucleon system \cite{BLS}, formed in a heavy ion collision, then
under lowering density this disassembling phase can undergo a first order
phase transition in density from the normal to superfluid state.

The temperature dependence of the energy gap was studied in
Refs.~\cite{SAL,SL}
with the use of the bare interaction in the gap equation in the form of
the Paris and Argonne V$_{14}$ potentials, respectively. Our results  agree
qualitatively  with theirs at   $T=0$. However, in our calculations with the
effective density--dependent NN interaction we obtain the gap as a function
of density (not at fixed density) and this
allows us to find new important features in behavior of the energy gap.

In summary, we studied TS superfluidity of asymmetric nuclear matter in the FL
  model with density--dependent Skyrme effective interaction (SkP force). In
the FL approach the normal and anomalous FL amplitudes are taken into account
on an equal footing  and this allows us to consider consistently within the
framework of a phenomenological theory the influence of medium effects on
superfluid properties of nuclear matter. It is shown, that for some values of
density and asymmetry parameter of nuclear matter the critical temperature of
a second order phase transition demonstrates double--valued behavior, that
agrees with the results of previous studies. In the case when the chemical
potentials $\mu_{00}, \mu_{03}$ (half of a sum and half of a difference of
the proton and neutron chemical potentials, respectively) are determined in
the approximation of ideal Fermi--gas, the energy gap demonstrates
for some values of density and asymmetry parameter the
double--valued behavior. If we consider the feedback of
pairing correlations through the dependence of the normal distribution
functions of nucleons  from the energy gap,  then the energy
gap drastically changes its behavior from two--valued to universal
one--valued character. In spite of relative smallness of the ratio
$\Delta/\varepsilon_F$, taking  into account of the finite size of the gap in
chemical potentials leads to the significant increase of absolute value of
$\mu_{03}$ and, hence, to considerable reduction of the threshold asymmetry,
at which superfluidity at $T=0$ disappears. In self--consistent determination
the energy gap at $T=0$ as a function of density has a finite width and
normal--to--superfluid and superfluid--to--normal phase transitions should
appear as a first order phase transitions in
density.  Among the other problems we note here the study of multi--gap
superfluidity \cite{AIP2} in asymmetric nuclear matter.

{\bf Acknowledgement.} The authors thank  A. Sedrakian for
reading the preliminary version of the manuscript and  valuable comment.
Financial support of
BMBF and STCU (grant $\#1480$) is  acknowledged.

\newpage
\begin{figure}[htbp]
\caption{Critical temperature as a function of density.
}\label{fig1}\end{figure}
\begin{figure}[htbp]
\caption{Energy gap  as a function of density in uncoupled calculations.
}\label{fig2}\end{figure}
\begin{figure}[htbp]
\caption{Energy gap  as a function of density in a self--consistent scheme.
}\label{fig3}\end{figure}
\begin{table}\caption{The values of the boundary points (in fm$^{-3}$),
determining the intervals of double-- and one--valued behavior of the energy
gap in uncoupled calculations. \label{table1}}\begin{tabular}{ddddd}
$\alpha$&$\varrho_{min}$&$\varrho_{min}'$&$\varrho_{max}'$&$\varrho_{max}$\\
\tableline 0.06&0.0004&0.0045&0.070&0.121\\
0.07&0.0008&0.016&0.035&0.116\\
0.072&0.0017&--&--&0.113\\
0.09&0.0018&--&--&0.102\\
0.14&0.0053&--&--&0.068\\
\end{tabular}
\end{table}
\end{document}